# Switching Current Measurements in Josephson Rings

K. Segall

*Department of Physics and Astronomy, Colgate University, Hamilton, NY 13346*

315-228-6597 (ph), 315-228-7038 (fax), ksegall@mail.colgate.edu

A. Dioguardi and N. Fernandes

*Department of Physics and Astronomy, Colgate University, Hamilton, NY 13346*

J. J. Mazo

*Dpto. de Física de la Materia Condensada, Universidad de Zaragoza and Instituto de Ciencia de Materiales de Aragon, C.S.I.C.-Universidad de Zaragoza, 50009 Zaragoza, Spain*

**Abstract:** We present switching current measurements in niobium-aluminum oxide-niobium underdamped Josephson rings. Underdamped Josephson rings display hysteretic current-voltage curves which can be characterized by their switching current, the value of applied current at which the junctions switch to the energy-gap voltage. The value of the switching current is strongly affected by temperature and by the presence of fluxons in the system. We observe a very small voltage (~ 0.2 V) across the ring prior to switching, indicating a low-velocity fluxon diffusion state before the jump to a full running state. In analogy with previous work on single junctions, we analyze the switching current data with the process of thermal activation over a dissipation barrier, where the system switches from a low velocity state to a high velocity state. We find that our data agrees qualitatively with this description, further supporting the observation of fluxon diffusion.

# I. Introduction

Arrays of Josephson junctions have proven to be extremely interesting systems to study nonlinear and collective effects [1]. Previous work has focused mostly on three major geometries: the two-dimensional array [2], the Josephson ladder [3], and the parallel array [4]. Of these the parallel array (or Josephson ring if it is closed in a circle), displays one-dimensional behavior and is an important component in RSFQ (Rapid-Single-Flux-Quantum) circuits. The main excitation in a Josephson ring is a fluxon, a discrete counterpart to the soliton. Fluxons in a Josephson ring obey the discrete Sine-Gordon equation and are analogous to kinks in the 1-D Frenkel-Kontorova model [5].

In overdamped Josephson rings, the current-voltage (I-V) curves are single valued. The dynamics of fluxons trapped in the ring are straightforward to infer in those cases because the voltage, which is directly proportional to the fluxon velocity, has a unique value at each current. Underdamped arrays display a richer variety of behavior but are more difficult to analyze, since the I-V curves are bistable. In many previous experiments on single junctions, *switching current* measurements have been utilized to study the dynamics of the Josephson phase when the I-V curves are bistable [6-10]. In these measurements, the value of the applied current at the point where the junction switches into the voltage state is recorded many times at each temperature. Switching current measurements have been used to infer thermal activation [6-7], quantum tunneling [7-8], phase diffusion [9], and to infer the state of a quantum bit [10].

In this paper we present switching current measurements on Josephson rings. First, the current-voltage (I-V) characteristics are presented and discussed. We find that a very small voltage appears across the ring prior to switching, indicating a low-velocity fluxon diffusion state. We also find that the switching current is different under positive and negative current, which we ascribe to the asymmetric manner in which the current is applied. Finally, the dependence of the switching current histograms on temperature is presented and analyzed in detail. We find that we can qualitatively fit the switching rate curves with the process of thermal activation over a dissipation barrier, where a low-velocity state switches to a high velocity state.

## II. Measurements

### A. Device and measurements

The device studied was a circular array of $N=9$ niobium-aluminum oxide-niobium Josephson junctions, fabricated at M.I.T. Lincoln Laboratory at a current density of 300 A/cm$^2$. A scanning electron micrograph of a device similar to the one tested is shown in Figure 1a. Each junction is about 1.05 μm on a side (area of about 1.1 μm$^2$) with a critical current of about $I_{crit} = 3.3$ μA and a normal state resistance of about $R_N = 678$ Ω. The circular array is about 133 μm in diameter with a total area of about $A_{tot} = 13{,}900$ μm$^2$. The inductance of the 9 individual cells was calculated with FASTHENRY [11] and found to be about $L = 79$ pH. The normalized damping $\eta = \sqrt{\Phi_0/\left(2\pi I_{crit} R_N^2 C_J\right)}$ is about 0.06 and the normalized discreteness parameter $\lambda = \Phi_0/(2\pi L I_{crit})$ is about 1.2; here $\Phi_0$ is the flux quantum and $C_J$ is the junction capacitance.

The devices were cooled in a $^3$He refrigerator and measured at temperatures of about 0.3 K to about 6 K. The voltage and current leads were filtered with copper powder filters and RC low-pass filters. The device was enclosed in a copper can which was heat sunk to the $^3$He stage. Fluxons were trapped in the ring upon cooling through the superconducting critical temperature [12]. A saw-tooth current wave at 100 Hz was applied to the array through a 100 kΩ bias resistor located at room temperature. In series with each junction is a 10 Ω on-chip bias resistor. Figure 1b shows the way in which the current was applied to and extracted from the device. Notice the asymmetry: current enters the device through nine separate wires, one for each junction, while current leaves the device through just one wire. This has important consequences for fluxon motion and is discussed below. The voltage across the array was measured with an instrumentation amplifier. The voltage and current waves were digitized directly and saved on disk. The switching currents were determined with software routines. Typically 5000 I-V curves were digitized at each temperature.

**B. Current-Voltage (I-V) curves**

Individual I-V curves at T = 0.35 K and T = 3.00 K are shown in Figure 2. These I-V curves look very similar to the I-V curve for a single underdamped Josephson junction. Current increases from zero along an apparent supercurrent branch and then the system switches to a voltage of about 2.8 mV, the energy gap for niobium. After the current is decreased to close to zero, the system retraps back to zero voltage. The switching current is defined as the current at which the jump to the energy gap voltage occurs. As seen in the figure, lower temperatures result in larger switching currents.

In Figure 2 one can also see that at a given temperature, the switching current is larger for negative currents than for positive currents. This was unexpected, since the junctions all have nominally the same critical current and the cells all have the same inductance. We ascribe this asymmetry to the way that we apply and extract the current. The return current comes through a single wire, as shown in Figure 1, and this one wire carries the bias current for all nine junctions. The magnetic field from this return current puts flux of one sign in the cell to the left and flux of the other sign in the cell to the right. This flux affects the motion of the fluxon, deepening the potential well on one side and raising it on the other. We have modeled this numerically with the equations of motion for the array and find that, qualitatively, it can cause an asymmetry in the switching current [13]. We have removed this asymmetry in a new set of devices, and preliminary results show equal switching current for currents of either sign.

If we consider a Josephson ring with a fluxon trapped inside, the dynamics one would typically infer from the I-V curves of Figure 2 are similar to that of the phase particle for a single junction. On the supercurrent branch, the fluxon is pinned in its potential minima. The energy barrier to move it to the next cell is lowered under the application of the applied current, and at the switching current the fluxon escapes and begins to move through the array, causing a voltage. The escape is triggered by thermal fluctuations in the system. Because the system is underdamped, the moving fluxon excites a whirling mode, and all the junctions switch to the energy gap voltage. The hysteresis results because in order to retrap the fluxon, the current needs to be reduced almost to zero.

However, this picture is not exactly correct in our device. In Figure 3 we show I-V curves at 0.35 K, 1.25 K and 3.00 K, but with the voltage axis expanded significantly. Here the I-V curves have been averaged 5000 times to reduce electronic noise. Now we can see that the apparent supercurrent branch is actually at a very small voltage, of order 200 nV. This is an important observation, because it indicates that the fluxon escapes not from a stationary state, but rather from a dynamic state.

We have recently explored the physics of this low-voltage state in a computational study [14]. This state is characterized by a non-zero mean velocity of the fluxon, where it moves by a series of noise-induced phase slips. We term this motion fluxon diffusion, in analogy with phase diffusion in a single junction. Like phase diffusion, fluxon diffusion occurs in an underdamped, hysteretic system [15]. Unlike phase diffusion, however, fluxon diffusion does not need frequency-dependent damping to occur. In this paper we are not able to compare our experimental results directly to our numerical simulations due to the asymmetry in current application, described above. However, we can confirm the presence of fluxon diffusion in our data by the observation of the low voltage prior to switching (Figure 3) and the analysis of the switching data (see below).

**C. Switching current measurements**

For each I-V curve measured, a switching current was determined at a voltage threshold of (+1 mV) for positive current and (-1 mV) for negative current. The retrapping currents were computed in a similar way. At each temperature we measured 5000 switching currents and retrapping currents. Figure 4 shows the histograms for four different temperatures: 0.35 K, 1.25 K, 3.00 K and 5.18 K. Also shown is a histogram of the retrapping events at 3.00 K. Here, and for the other switching current data and analysis, we have presented results for only negative applied currents for brevity. Results for positive applied currents gave similar results as those for negative applied currents except with different fitting parameters, which we indicate when appropriate.

The width of the retrapping histogram in Figure 4 is extremely narrow, as the process of retrapping has intrinsically much smaller fluctuations than the process of switching. We have not yet explored the physics of the retrapping process in our devices, but for now we use it as a measure of experimental uncertainty due to electrical noise or other external sources. From the narrow width of the retrapping histogram in Figure 4 we conclude that the experimental uncertainty is very small; we would not be able to measure such a narrow histogram otherwise.

The shapes of the switching histograms are reminiscent of thermally activated switching in single junctions, with slightly more events at currents below the peak of the histogram than above the peak. We convert the histogram data into a switching rate per unit time $\Gamma$, using the procedure first introduced by Fulton and Dunkleberger [4]. The experimental ramp rate was $(dI/dt) = 1.8 \times 10^5$ µA/s. Rate curves as a function of current are shown for 7 different temperatures (0.35 K, 0.55 K, 1.25 K, 3.00 K, 3.50 K, 5.18 K and 5.58 K) in Figure 5. They show the expected exponential dependence on current. Finally, in Figure 6 we show the average switching current, $\langle I_{sw} \rangle$, and standard deviation, $\sigma_{Isw}$, for each distribution as a function of temperature. These data will be used in analyzing the rate curves below.

## III. Analysis

In this section we analyze the switching current data and its temperature dependence. We are specifically interested in understanding the rate curves from Figure 5 with the process of switching from a low-velocity state to a high-velocity state. We follow the work of Vion et al. [16], who study such a process in a Josephson system. Their device is a single Josephson junction of critical current $I_0$ biased in series with an on-chip resistor ($R$) and in parallel with an on-chip capacitance ($C$). At low bias currents, the junction is in a low-velocity, phase-diffusion state. At the switching current, the junction is thermally excited from this diffusion state to the high-voltage running state. In order to reach that voltage, the external *RC* circuit must be charged and energy must be dissipated; thus a "dissipation barrier" separates the two states.

After normalizing the equations of motion, the dynamics of their system are mapped to an effective Langevin equation. The rate Γ to switch to the running state is given by a static, over-damped Kramers rate. It is specified by equations (3)-(5) in Vion et al. [16] and involves various sums over modified Bessel functions. For ease of comprehension, we adopt the following empirical expression for Γ using the equations in Vion et al.:

$$\Gamma = \left(\frac{11\omega_J}{2\pi\alpha}\right)\left(1 - \frac{I}{I_{\max}(T)}\right)^{1/2} \exp\left[-\frac{\alpha E_J}{6kT}\left(1 - \frac{I}{I_{\max}(T)}\right)^{3/2}\right]. \quad (1)$$

Equation (1) was obtained by computing the value of Γ at many different values of current and temperature, and then fitting the current and temperature dependence. Here $\omega_J = 2\pi R I_0/\Phi_0$ is the junction frequency, $\alpha = 2\pi R^2 I_0 C/\Phi_0$ is the dissipation factor, $E_j = \Phi_0 I_0/(2\pi)$ is the Josephson energy, *k* is Boltzmann's constant and *T* is temperature. $I_{max}(T)$ is the temperature-dependent maximum current for which the dissipation barrier still exists, equal to max[*S*(*u*)] in Vion et al. [16]; currents greater than $I_{max}(T)$ result in definite switching. The exponent in (1) matches the expressions in Vion et al. to 5% or better in the range of switching currents and temperatures studied. The pre-factor does not match quite as well (about 50% or better), but the pre-factor does not affect our results significantly.

Writing the rate in the form of equation (1) allows us to now use the general formalism developed by Garg [17]. Here the rate Γ is defined as follows:

$$\Gamma = A\left(1 - \frac{I}{I_c}\right)^{a+b-1} \exp\left[-B\left(1 - \frac{I}{I_c}\right)^b\right]. \quad (2)$$

Equation (2) is a general expression that applies to switching in a Josephson junction with critical current $I_c$. $A$, $B$, $a$ and $b$ are constants which depend on temperature and damping. It is easily seen that equations (1) and (2) are equivalent for $I_c = I_{max}$, $a = 0$, $b = 1.5$, $A = 11\omega_j/(2\pi\alpha)$ and $B = \alpha E_j/(6kT)$. Equations (7) and (8) in Garg [17] also gives expressions for the average, $<I_{sw}>$, and standard deviation, $\sigma_{Isw}$, of a switching current distribution in terms of $a$, $b$, $A$, $B$ and $I_c$. To plot equation (2) in a manner that is equivalent for all temperatures, we define the rescaled current as:

$$Z = B\left(1 - \frac{I}{I_c}\right)^b, \qquad (3)$$

and the rescaled rate as:

$$F = \ln\left[\frac{\Gamma}{A(1 - I/I_c)^{a+b-1}}\right]. \qquad (4)$$

Using these definitions, (2) can then be written as:

$$F = -Z. \qquad (5)$$

Thus, curves for different temperatures should all fall onto the same curve defined in (5) provided the correct values of $a$, $b$, $A$, $B$ and $I_c$ have been used at each temperature.

We analyze the rate curves in figure 5 by first finding the Garg parameters $B$ and $I_c$ that scale all of the curves onto the form given by equation (5), and then comparing them to what we expect from equation (1). First, as suggested by equation (1), we choose $b = 1.5$, $a = 0$ and hold $A$ at a constant value for all temperatures. Then, we use the measured values of $<I_{sw}>$ and $\sigma_{Isw}$ at each temperature (Figure 6) to choose values for $B$ and $I_c$. This is done by numerically back-solving for $B$ and $I_c$ using the expressions in Garg [17] for $<I_{sw}>$ and $\sigma_{Isw}$.

Finally, this step is repeated for different values of $A$ until all curves collapse onto the form given by (5). Effectively what we have done is to use equation (2) to convert our measurements of $<I_{sw}>$ and $\sigma_{Isw}$ into measurements of $B$ and $I_c$, which are parameters that have more physical meaning.

The results are shown in Figures 7 and 8. In Figure 7, we show the results of the scaling. All seven temperatures shown in Figure 5 now collapse onto the same curve. The values of $I_c$ and $B$ needed to obtain this scaling are shown in Figure 8. The value of $A$ used was $4 \times 10^8$ Hz. As best as we could tell, no other set of parameters scaled the curves in a similar fashion.

We now interpret these fitted values of the parameters in terms of switching from one dynamical state to another. Incidentally, it is clear immediately that thermal activation or quantum tunneling from a stationary state do not describe our system, since the fitted $I_c$ depends on temperature. (This is also clear since the standard deviation does not increase with increasing temperature.) In switching from a dynamic state, however, the maximum current $I_{max}$ does depend on temperature, since thermal fluctuations affect the height and shape of the dissipation barrier. In Figure 9 we compare the temperature dependence of the fitted critical current with the temperature dependence of $I_{max}$, calculated using the expressions in Vion et al. Only one fitting parameter, $I_0$, is used. This parameter scales our applied current as $I/(NI_0)$ and also scales our temperature as $kT/E_j$, since $E_j$ is proportional to $I_0$. We see that the temperature dependence of our fitted $I_c$ is consistent with $I_{max}(T)$ for a dissipation barrier. A fitted value of $I_0 = 1.6$ µA was used in Figure 9. For positive applied current, a value of $I_0 = 1.05$ µA best fits the data.

We next look at the fitted values of the $B$ coefficient. In Figure 10 we plot $B$ versus $E_j/(kT)$. A linear relationship is found for values of $E_j/(kT)$ between about 15 and 40. Using $B = \alpha E_j/(6kT)$ from equation (1), we find that $\alpha = 30$ fits the data reasonably well. Deviations from the expected relationship occur at both low temperature and high temperatures. At high temperatures, or low values of $E_j/(kT)$, the retrapping current of the array increases significantly, indicating increased dissipation in the array junctions themselves. This causes the value of $\alpha$ to increase at these temperatures, resulting in the larger values of $B$. At low temperatures, the value of $B$ levels off and is smaller than the predicted value. It is worth pointing out that this is the prediction of quantum tunneling, although many more measurements and justification would be needed to make this claim.

The other fitting parameter, $A$, has a value of $A = 4 \times 10^8$ Hz. Using $A = 11\omega_j/(2\pi\alpha)$ and $\alpha = 30$ from above gives a fitted junction frequency of about $\omega_j = 7$ GHz; this then corresponds to a value of $R = 1.2$ $\Omega$ and $C = 3.15$ nF for the dissipation barrier of the fluxon. The fitting parameters are discussed further below.

## IV. Discussion

We have performed measurements of switching current distributions on 9-junction Josephson rings at low temperatures. The rings are highly underdamped, resulting in bistability in the I-V curves. A small voltage, of order hundreds of nanovolts, is measured across the junction before the jump to the voltage state. Histograms of switching current data are qualitatively consistent with the process of thermal activation over a dissipation barrier as described by Vion et al. [16], where a low-velocity state switches to a high-velocity state. These two observations both indicate the presence of low-velocity fluxon diffusion before switching. To the best of our knowledge, fluxon diffusion in an underdamped, hysteretic system has not been observed previously.

In using the model of Vion et al. to fit our data we are mapping the dynamics of the fluxon in our ring, which is inherently a collective excitation, onto the dynamics of a single particle, that being the phase particle of a single junction. Following that analogy, we expect the fitted critical current $I_0$ to be equal to the zero-temperature switching current of the fluxon (the current where the fluxon excites a whirling mode). Unfortunately it is difficult to determine exactly what this current is, since we do not have the ability to accurately model the extra flux due to the return lead (see above). However, we can say that $I_0$ should not be smaller than the fluxon depinning current (about 0.13 µA) and should not be larger than the critical current of the junctions in the ring ($I_{crit}$ = 3.3 µA), so the fitted values of 1.05 µA and 1.6 µA do satisfy those restrictions.

The dissipation barrier for the fluxon is set by the bias resistors, the frequency-dependent damping due to the bias leads, and the radiation damping of the fluxon; this is to be compared to the external RC circuit in the Vion et al. experiment. The fitted value of $R$ = 1.4 Ω corresponds very close to parallel-equivalent of the 9 bias resistors (1.1 Ω). The fitted value of $C$ = 3.13 nF is larger than one would expect for the capacitance to ground of our biasing leads. This indicates that perhaps the radiation damping of the fluxon and the frequency-dependent damping due to the bias leads contribute significantly to the effective dissipation barrier. Our future experiments and simulations will focus on studying this issue.

Acknowledgements

Work is partially supported by DGICYT through project FIS2005-00337 (J.J.M) and through NSF DMR 0509450 (K.S.)

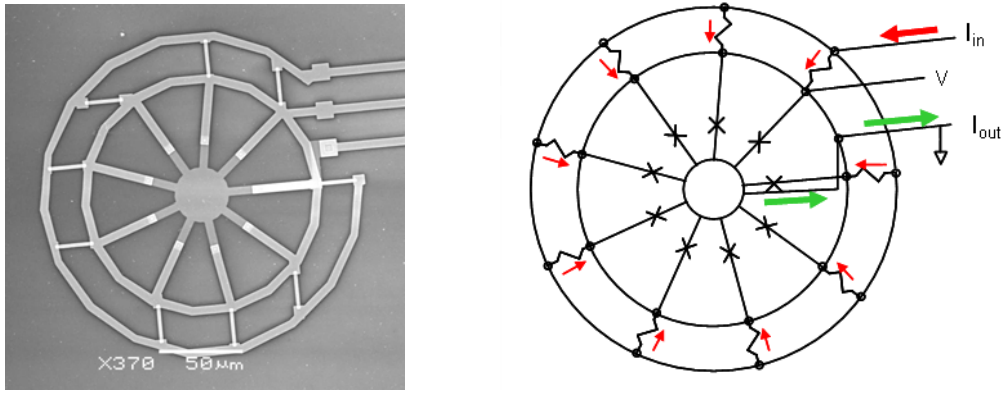

Figure 1: Scanning electron micrograph (a) and schematic of the device (b). In the schematic, the junctions are indicated with X's and the application of the current is shown. Red arrows indicate current entering the device while green arrows indicate current leaving the device.

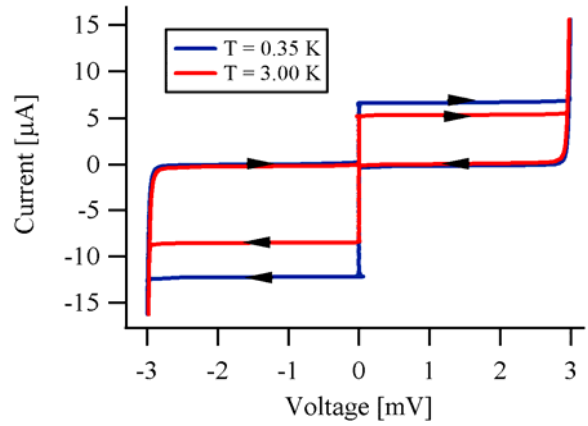

Figure 2: Single I-V trace for the Josephson ring at 0.35 K (blue) and 3.00 K (red).

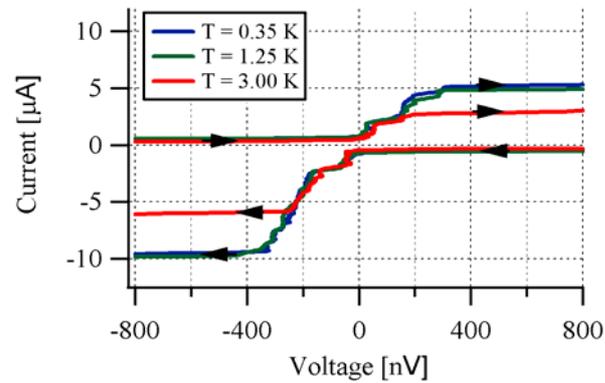

Figure 3: Zoom-in of the average I-V curve for 0.35 K, 1.25 K, and 3.00 K. The voltage across the array is in the 200-300 nV range before switching and indicates the presence of low-velocity fluxon motion.

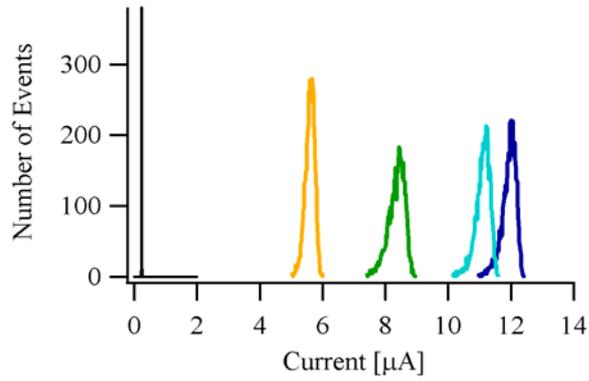

Figure 4: Histograms of the retrapping current at 3.00 K (black) and for the switching current at 0.35 K (purple), 1.25 K (blue), 3.00 K (green), and 5.18 K (yellow).

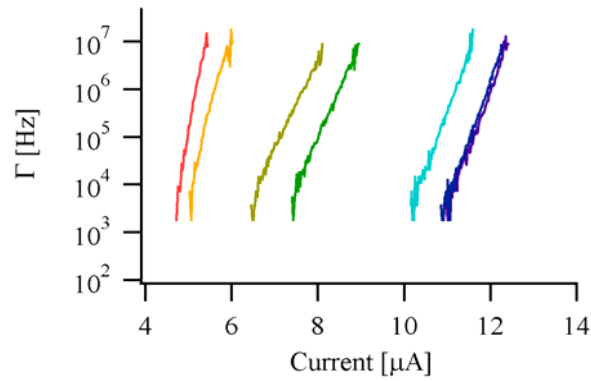

Figure 5: Switching rate ($\Gamma$) versus applied current for 7 different temperatures. The temperatures are, from right to left: 0.35 K, 0.55 K, 1.25 K, 3.00 K, 3.50 K, 5.18 K, 5.58 K. The rate shows an exponential dependence on current.

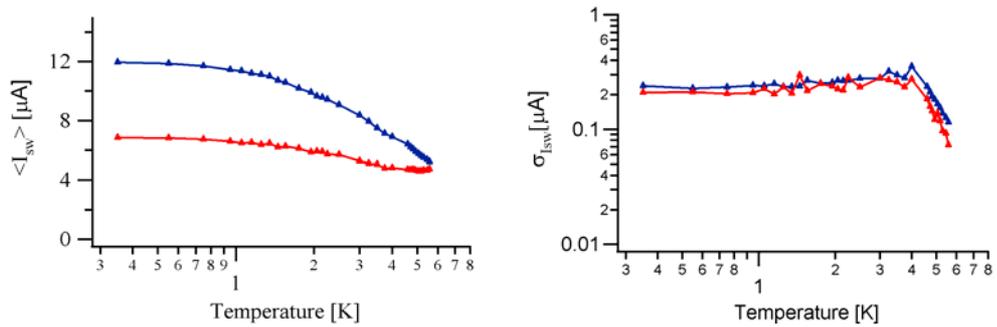

Figure 6: Switching current average (a) and standard deviation (b) as a function of temperature. Blue indicates the minus direction and red indicates the plus direction.

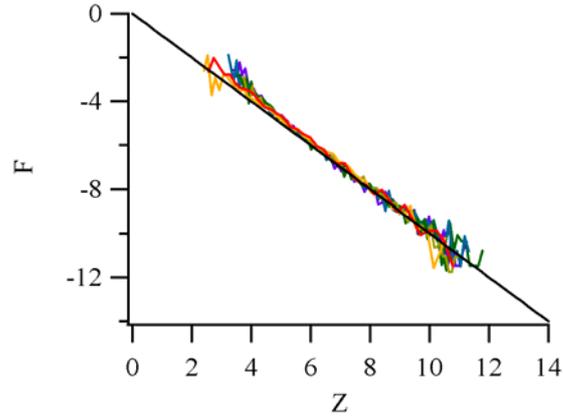

Figure 7: Normalized rate (*F*) versus normalize current (*Z*) for the same seven rate curves in Figure 5. Color coding is the same as in Figure 5.

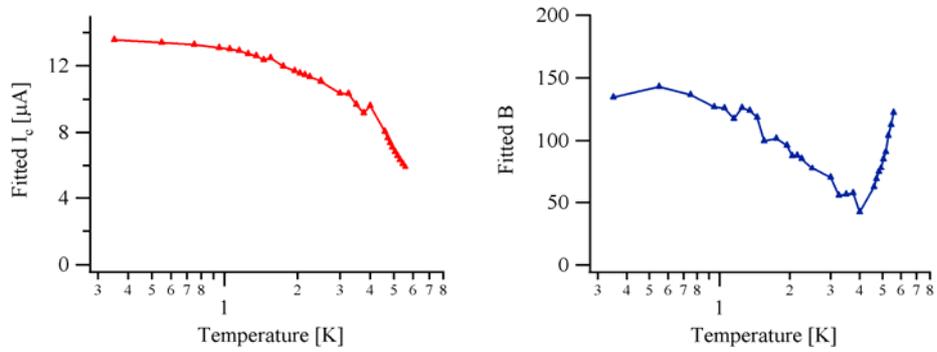

Figure 8: Fitted $I_c$ (a) and *B* (b) versus temperature. The lines are a guide to the eye.

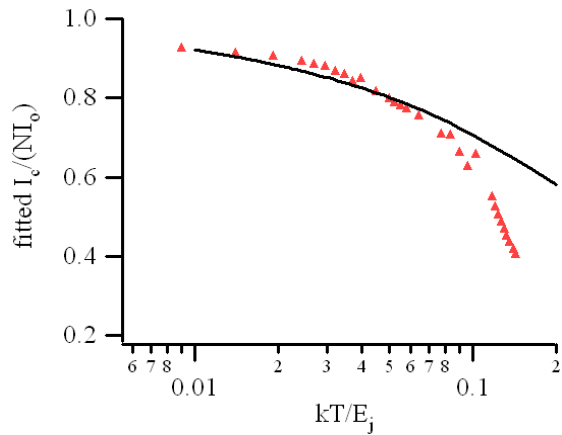

Figure 9: Fitted $I_c$ normalized to $NI_o$ (triangles) versus normalized temperature ($kT/E_j$). Black line is the curve $I_{max}(T)$ calculated from Vion et al.

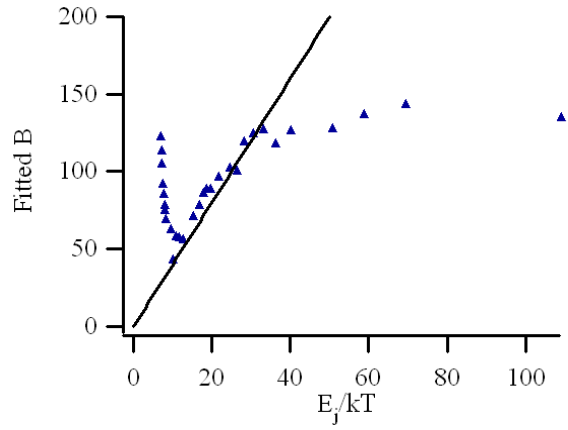

Figure 10: Fitted *B* (triangles) versus $E_j/kT$. The black line is a fit with $\alpha = 30$.